\documentclass[preprint]{aastex}

\begin{document}

\title{Long-term Photometric Behavior of the Eclipsing Binary GW Cephei}
\author{Jae Woo Lee, Jae-Hyuck Youn, Wonyong Han, Chung-Uk Lee, Seung-Lee Kim, Ho-Il Kim, Jang-Ho Park$^{1}$, and Robert H. Koch$^2$}
\affil{$^1$Korea Astronomy and Space Science Institute, Daejeon 305-348, Korea}
\email{jwlee@kasi.re.kr, jhyoon@kasi.re.kr, whan@kasi.re.kr, leecu@kasi.re.kr, slkim@kasi.re.kr, hikim@kasi.re.kr, pooh107162@kasi.re.kr}
\affil{$^2$Department of Physics and Astronomy, University of Pennsylvania, Philadelphia, USA}
\email{rhkoch@earthlink.net}

\begin{abstract}
New CCD photometry during 4 successive years from 2005 is presented for the eclipsing binary GW Cep, together with 
reasonable explanations for the light and period variations. All historical light curves, obtained over a 30-year interval, 
display striking light changes, and are best modeled by the simultaneous existence of a cool spot and a hot spot on 
the more massive cool component star. The facts that the system is magnetically active and that the hot spot has consistently 
existed on the inner hemisphere of the star indicate that the two spots are formed by (1) magnetic dynamo-related activity 
on the cool star and (2) mass transfer from the primary to the secondary component. Based on 38 light-curve timings 
from the Wilson-Devinney code and all other minimum epochs, a period study of GW Cep reveals that the orbital period has 
experienced a sinusoidal variation with a period and semi-amplitude of 32.6 yrs and 0.009 d, respectively. In principle, 
these may be produced either by a light-travel-time effect due to a third body or by an active magnetic cycle of at least 
one component star. Because we failed to find any connection between luminosity variability and the period change, 
that change most likely arises from the existence of an unseen third companion star with a minimum mass of 0.22 $M_\odot$ 
gravitationally bound to the eclipsing pair.
\end{abstract}

{}

\section{INTRODUCTION}

After GW Cep (CSV 5941, BV 7, 2MASS J01455862+8004553) was discovered to be a variable star by Strohmeier (Geyer et al. 1955), 
its light curves were made photoelectrically by Meinunger \& Wenzel (1965), Hoffmann (1982), Landolt (1992), and 
Pribulla et al. (2001). They recognized it as a W-subtype (defined empirically by Binnendijk (1970)) eclipsing binary 
with complete eclipses. Kaluzny (1984) analyzed Hoffmann's $BV$ light curves by using Rucinski's (1976) code and concluded 
that the binary is an over-contact system with the characteristic parameters of $q$=2.703 and $i$=83$^\circ.9$ and 
a fill-out factor $f$=11.3 \%. Pribulla et al. analyzed their own $BV$ light curves using 
the Wilson \& Devinney (1971, hereafter W-D) code. Except for a value of $f$=23.5 \%, their results are very similar to those 
of the earlier study. The orbital period of GW Cep has been examined by Pribulla et al. and Qian (2003), who reported that 
the period is decreasing. More recently, Chochol et al. (2006) suggested that the variation of the orbital period could be 
produced by two possible forms: a single light-travel time (LTT) ephemeris with and without a quadratic term.

Although GW Cep has been studied photometrically these several times, intrinsic light variations due to starspots have 
not yet been considered. Additionally, the period variation still has not been described as conclusively as can be desired. 
In this paper, we present and discuss the long-term photometric behavior of the binary system from detailed studies 
of all available data.

\section{NEW EXTENDED CCD PHOTOMETRY}

We carried out CCD photometric observations of GW Cep on 20 nights from 2005 October through 2008 November in order to 
look for possible long-term light variability. The observations were taken with a SITe 2K CCD camera and a $BVR$ filter set 
attached to the 61-cm reflector at Sobaeksan Optical Astronomy Observatory (SOAO) in Korea. The exposure times were 
about 40$-$85 s for $B$, 20$-$40 s for $V$, and 10$-$20 s for $R$ with a 2$\times$2 binning mode. The individual choices 
depended on 
the seeing and transparency of the night sky. The instrument and reduction method have been described by Lee et al. (2007a). 
A summary of the observations is given in Table 1, where we present seasonal observing intervals, filters, numbers of 
observed points, and designations for all datasets. 

An image of the observing field appears in Figure 1, with the eclipsing variable designated as GW and candidate comparison stars 
marked as Rn (n being a sequence number). Since an individual frame was large enough to image a few tens of nearby stars 
simultaneously, we monitored many of them on each frame. To find a comparison star that would be optimal 
for long-term observing, we first took an average of 5 potentially useful field stars calling that average $<$R$>$ and 
examined each field star against $<$R$>$ as a reference. The R1 and R3 stars showed excessive noise and were therefore 
excluded as possible comparison candidates. Candidates R4 and R5 did not appear to be variable but their noise levels 
against $<$R$>$ were larger than was the case for R2.  Ultimately, R2 (2MASS J01405717+8011069, GSC 4502-0542, TYC 4502-542-1;
$V\rm_T$=$+$10.82, $(B-V)\rm_T$=$+$0.94) was chosen as a suitable comparison star (C). The 1$\sigma$-value of the dispersion 
of the (C$-<$R$>$) differences is about $\pm$0.006 mag. 

A total of 4,227 individual observations was obtained among the three bandpasses (1414 in $B$, 1416 in $V$, and 1397 in $R$) 
and a sample of them is listed in Table 2.  The light curves are plotted in Figure 2 as differential magnitudes 
{\it versus} orbital phase.

\section{THE LIGHT CURVES AND SPOTS MODELS}

Figure 2 displays a quite active light curve for GW Cep. Most observations show the conventional O'Connell effect with Max I 
(at phase 0.25) brighter than Max II.  Our light curves were solved in a manner similar to that for TU Boo (Lee et al. 2007b) 
and AR Boo (Lee et al. 2009a) by using contact mode 3 of the W-D code. In order to obtain an unique set of photometric solutions, 
our light-curve synthesis was carried out in two stages. In the first stage, all SOAO observations were analyzed simultaneously 
without a spot. In the second stage, the existence of short-time brightness disturbances caused us to model separately each dataset 
(designated in the fifth column of Table 1) by assuming the unspotted solutions and by adjusting only spot and luminosity parameters. 

The surface temperature of the more massive star was held fixed at 5800 K, appropriate for its spectral type G3 
(Meinunger \& Wenzel 1965). Linear bolometric ($X$) and monochromatic ($x$) limb-darkening coefficients were initialized 
from the values of van Hamme (1993) and were used together with the model atmosphere option. From a detailed $q$-search procedure 
over the range of 0.3 $\le q \le$ 4.0, we found that the sum of the weighted squared residuals, $\Sigma W(O-C)^2$, 
reached a minimum around $q$=2.60.  That indicates that GW Cep is indeed a W-subtype contact binary as previous workers had found. 
The initial value of $q$ was then treated as an adjustable parameter to derive the unspotted solution listed in Table 3, 
wherein the primary and secondary stars refer to those being eclipsed at Min I and Min II, respectively. Therefore, the latter star is 
the cooler, larger, and more massive component. The $V$ residuals from the analysis are plotted in the left panels of Figure 3.  
Similar patterns exist for the other bandpasses.

As can be seen from these panels, the model light curves do not fit the observed ones at all well. The discrepancies resemble 
those of solar-type contact binaries recently studied such as TU Boo, AR Boo, and BX Peg (Lee et al. 2009b), for which 
non-modelled light could be explained by spot variability on the stellar photospheres.  This condition could result from 
magnetic dynamo-related activity and/or variable mass transfer between the components. Because our long-term data show 
conspicuous seasonal light variations despite a constant value of $f$, it is more reasonable to regard the main cause of 
the second-order activity to be a magnetized cool spot on either component star of GW Cep rather than 
sporadically-variable mass transfer. Adopting the light curves of SOAO6 as reference ones, we tested a single cool spot 
on each component. A cool spot on the larger secondary star gives a slightly smaller value of $\Sigma W(O-C)^2$ than 
if the spot were on the primary. This is consistent with Mullan's (1975) suggestion that the more massive components of 
contact binaries would preferentially manifest starspots. Therefore, we solved  each SOAO dataset separately using 
the cool-spot model. The results are given as the one-spot model of Table 4 and the light residuals from this model are 
plotted in the central panels of Figure 3. It shows that a single cool spot on the secondary component does 
improve the light-curve fitting greatly but brightness disturbances are still conspicuous around phase 0.38. This detail 
can be interpreted by introducing a hot spot on the surface of the secondary as was verified by testing the SOAO6 data 
for this possibility. Lastly, all SOAO light curves were re-analyzed by using a two-spot model with both a cool spot and a hot spot 
on the secondary star. The final result is listed in the two-spot-model entries of Table 4 and the residuals from this model 
are plotted in the right panels of Figure 3. Although not perfectly flat with respect to phase, this distribution of residuals 
is a great improvement over the previous one.

To study the long-term spot behavior of GW Cep, we analyzed the historical light curves by using our photometric parameters 
listed in Table 3 and adopting a two-spot model for the more massive secondary component. The data of Meinunger \& Wenzel (1965) 
could not be included in our analysis because they remain unavailable. The model parameters for the datasets of Hoffmann (1982), 
Landolt (1992), and Pribulla et al. (2001) lead to the continuous curves in Figures 4--6 and appear in Table 5. 
Although Landolt's observations have a large gap between phases 0.37 and 0.68, his light curves were actually modeled for 
spot descriptions. It can be seen that the stellar brightness ratios have not changed as a result of the spot modeling and 
it can also be seen that, in the 2000 data by Pribulla et al., the cool spot had almost disappeared. From the analyses, 
we conclude that the two-spot model satisfies all GW Cep curves quite well and gives a good representation of the binary system 
for both the photospheric and spot descriptions. As seen in Tables 4 and 5 and Figure 7, our results indicate that 
the long-term light variations are dominated by the changes of the cool spot with time, especially by a longitude drift of 
that spot. Because its position on the inner hemisphere of the secondary star has not changed appreciably over the 30-year interval, 
the hot spot can be produced by stable mass transfer from the less massive primary impacting on the more massive secondary component. 
Thus, the intrinsic light variations of GW Cep have been modeled by the simultaneous existence of the active cool spot on 
the secondary star and an enduring hot spot due to mass transfer between the components. In all the procedures that have been 
described, we did not look for a possible third light source as might be suggested by the orbital period study now to be described. 

Identifying a satisfactory comparison star is critical to the credibility of the spot descriptions. Had we, for instance, 
chosen R1 as the comparison star, a light modulation of a bit less than 0.02-mag peak-to-peak would have been introduced 
into the sequence of light curves and this could well have been attributed to a third spot with a cycle length of about 3 years.

\section{ORBITAL PERIOD STUDY}

Times of minimum light may be shifted from conjunction instants by asymmetrical eclipse minima due to spot activity 
(cf. Lee et al. 2009a). Because 9 datasets of GW Cep were modeled for spot parameters, we calculated a minimum epoch 
for each eclipse in these datasets with the W-D code by means of adjusting only the ephemeris epoch ($T_0$). 
Thirty-eight such timings of minimum light are given in Table 6, together with those previously known or newly determined 
using the method of Kwee \& van Woerden (1956, hereafter KvW) for comparison. Two of these (HJD 2,444,289.27664 and 
2,451,884.61130) have been derived by us from the individual measurements. As in the case of AR Boo,
there are systematic runs of differences between the KvW timings and the W-D ones. For SOAO6 and probably for SOAO2-5 as well, 
the differences are negative for Min I and positive for Min II except for two eclipses (HJD 2,454,026.99936 and 2,454,421.07651).
These differences are caused by the cool spot and the hot spot presented to the observer before Min I and Min II, respectively. 
A total of 156 minimum timings over 45 years (50 visual, 6 photographic, 24 photoelectric, and 76 CCD), including 
the new W-D timings, were used to study the orbital period. For ephemeris computations, the following standard deviations 
were assigned to timing residuals based on the observational technique and the method of measuring the epochs: $\pm$0.0068 d 
for visual and photographic, $\pm$0.0012 d for photoelectric, $\pm$0.0008 d for CCD, and $\pm$0.0004 d for W-D minima.  
Relative weights were then scaled from the inverse squares of these values. 

From a parabolic least-squares fit, Pribulla et al. (2001) suggested that their measured period decrease can be driven by 
mass transfer from the more to the less massive component but may also be inflected by an LTT effect caused by 
the presence of a third companion physically bound to the eclipsing pair. The reason for this conclusion is that the sum 
of the weighted squared residuals from a 3rd-order polynomial fit is 7 \% lower than for a 2nd-order one.  
Chochol et al. (2006) reported that the orbital period changes of the binary system can be represented by either 
a single LTT ephemeris or by a combination of a downward parabola and an LTT (i.e., a quadratic {\it plus} LTT ephemeris). 
To find the best representation, We fitted all minimum epochs to several ephemeris forms including those already tried 
and found that the single LTT ephemeris gives a satisfactory representation without needing a quadratic term:
\begin{eqnarray}
C = T_0 + PE + \tau_3,
\end{eqnarray}
where $\tau_{3}$ is the LTT due to a third body (Irwin 1952, 1959) and includes five parameters ($a\sin i_3$, $e$, $\omega$, 
$n$, $T$). The Levenberg-Marquart technique (Press et al. 1992) was applied to solve for the parameters of the ephemeris. 
The result is plotted in Figure 8 and orbital parameters are summarized in Table 7, together with the third-body masses 
($M_3$) calculated for three different values of $i_3$. Absolute dimensions by Maceroni \& van't Veer (1996) have been 
used for these and subsequent calculations. The orbital eccentricity is not significant.

If the hypothetical third companion is a normal main-sequence star and its orbit is coplanar with the eclipsing pair 
($i_3$ = 85$^\circ.4$), the mass of the object is $M_3$ = 0.22 M$_\odot$ and its radius is calculated to be 
$R_3$ = 0.23 R$_\odot$ from the empirical mass-radius relation of Bayless \& Orosz (2006). These correspond to 
a spectral type of about M6-7 and a bolometric luminosity of $L_3$ = 0.004 L$_\odot$. This small value compared to 
the larger intrinsic variability due to variable spots is the reason no ``third light" was sought in our analyses.

Because there is no independent evidence to support the LTT hypothesis, one must consider the alternative:  
a period modulation due to a magnetic activity cycle, as initially proposed by Applegate (1992) and later modified 
by Lanza et al. (1998).  According to this model, a change in the distribution of angular momentum of 
a magnetically active star causes a change in the star's gravitational quadratic moment and hence forces a modulation 
of the orbital period. To obtain the model parameters for possible magnetic activity, we applied the period ($P_3$) 
and amplitude ($K$) to Applegate's formulae. The results from these calculations are presented in Table 8 and 
correspond to typical values for contact binaries. Here, the quantity $\Delta m_{\rm rms}$ denotes 
a bolometric magnitude difference calculated from the equation (4) of Kim et al. (1997). 

According to the prediction of the Applegate model, the brightness variation of a binary should vary in a pattern 
and sense similar to the period change seen in the $O$--$C$ diagram. In order to study this possibility, we measured 
the light levels at four special phases for the GW Cep light curves. The $\Delta V$ variations for only the SOAO data 
are plotted in the first through fourth panels of Figure 9 because the other datasets used different comparison stars
and could not be made comparable to the zero points of the SOAO light curves. The light level at Max II, 
plotted in the third panel, has brightened in tandem with the period increase over the interval of monitoring but 
the other light levels fail the test.  To look for any further connection between spot variability and 
orbital period changes, we examined the changes with time of the radius ($\rm R_c$) of the cool spot and of 
the luminosity ratio ($\rm L_c / L_h$) between the cool and hot spots from our two-spot model. These have been assembled 
in the fifth and sixth panels of Figure 9, respectively.  There clearly exists a variation in the size of the cool spot
but the changes in its parameters do not conform to Applegate's prediction. At present, we do not know what would be 
the appearance of a complete cycle for the cool spot because data are insufficient. It is also possible that 
magnetic activity is not  constant, as already suggested by us for AR Boo, but the weight of present evidence favors
a 3rd-body interpretation rather than a magnetic-cycle one for the period modulation.

This is the second in our series of contact binary studies in which we have used timings developed from the W-D code.
As Figure 8 in this study and Figure 2 accompanying the paper on AR Boo show, this method is superior to all others:  
if the ordinate scale of an $O$--$C$ diagram is chosen so as to make obvious the noise in visual or photographic timings, 
the noise in photoelectric and conventional CCD timings will also be detectable but that from W-D determinations will not 
be obvious to even detailed inspection of the printed page.  The weight of such timings will be 
the most powerful determinant in fitting the timing residuals. This understanding must become part of any program decision 
of whether simply to monitor minimum timings or to develop entire light curves whereby the bias due to photospheric spots 
may be removed from the timings. Only in this way will large and small secular and cyclical terms be evaluated to 
the highest accuracy and it be possible to decide if the cyclical terms are really periodic.

\section{SUMMARY AND DISCUSSION}

New multiband CCD light curves plus historical ones all display complete eclipses leading to system descriptions of 
high weight. The O'Connell effect and superposed long-term light variability are best explained by both a cool spot 
and a hot one on the more massive secondary star. The long-term variability has been ascribed to changes in the size 
of the cool spot. This spot may be produced by magnetic dynamo-related activity because the system is rotating rapidly 
and has a common convective envelope. The hot spot, on the other hand, is likely caused by impact from mass transfer 
between the components.

Over 45 years, the orbital period change of GW Cep has shown a periodic oscillation with a cycle length of 32.6 yr 
and a semi-amplitude of 0.009 d. In principle, this oscillation may be produced either by an LTT effect due to 
an unseen third star with a scaled mass of $M_3 \sin i_3$=0.22 $M_\odot$ or by an active magnetic cycle in 
the more massive secondary star. We indicate that the latter interpretation is the less likely of the two possibilities. 

With a relatively favorable light ratio and complete eclipses GW Cep clearly merits a radial velocity determination.
Those eventual results will add weight to understanding the entire class of contact binaries but is there anything 
about the binary which appears singular at the present time?  One item which attracts attention is the unfamiliar finding 
that the Keplerian period has been constant over 45 years.  This may be examined in some detail 
from The Atlas of $O$--$C$ Diagrams of Eclipsing Binary Stars (Kreiner et al. 2001). From that source we extracted 
the period histories of all cool contact binaries which have periods shorter than that of GW Cep. These systems number 
32 (1 A-type, 25 W-types and 6 not presently classified) and the shortest period is 0.221 d. In order to complete a sample 
for which GW Cep will have the median period, we next found the 32 contact binaries which show successively longer periods.  
These 32 systems are apportioned among 11 A-types, 17 W-types and 4 presently unclassified and the longest period is 0.365 d. 
Of this total sample of 65 pairs, 6 (AT Aqr, GW Cep, V906 Cyg, V743 Sgr, RZ UMi and BP Vel) appear at present to show 
constant periods.  The remainder show secularly variable periods of either algebraic sign upon which bounded oscillations 
are superposed in some cases.  Clearly, constant periods are an uncommon phenomenon in the sample and one may conjecture 
that they signify brief episodes of constant period behavior during themal oscillation evolutionary events. Of the 6 binaries, 
all but GW Cep have poor continuity in their period histories leaving that object as the best system at present to test 
such a possibility. Not only GW Cep, but the 5 other binaries merit continuing period monitoring.

\acknowledgments{ }
We would like to thank the staff of the Sobaeksan Optical Astronomy Observatory for assistance with our observations.
This research has made use of the Simbad database maintained at CDS, Strasbourg, France.

\clearpage
\begin{figure}
 \includegraphics[]{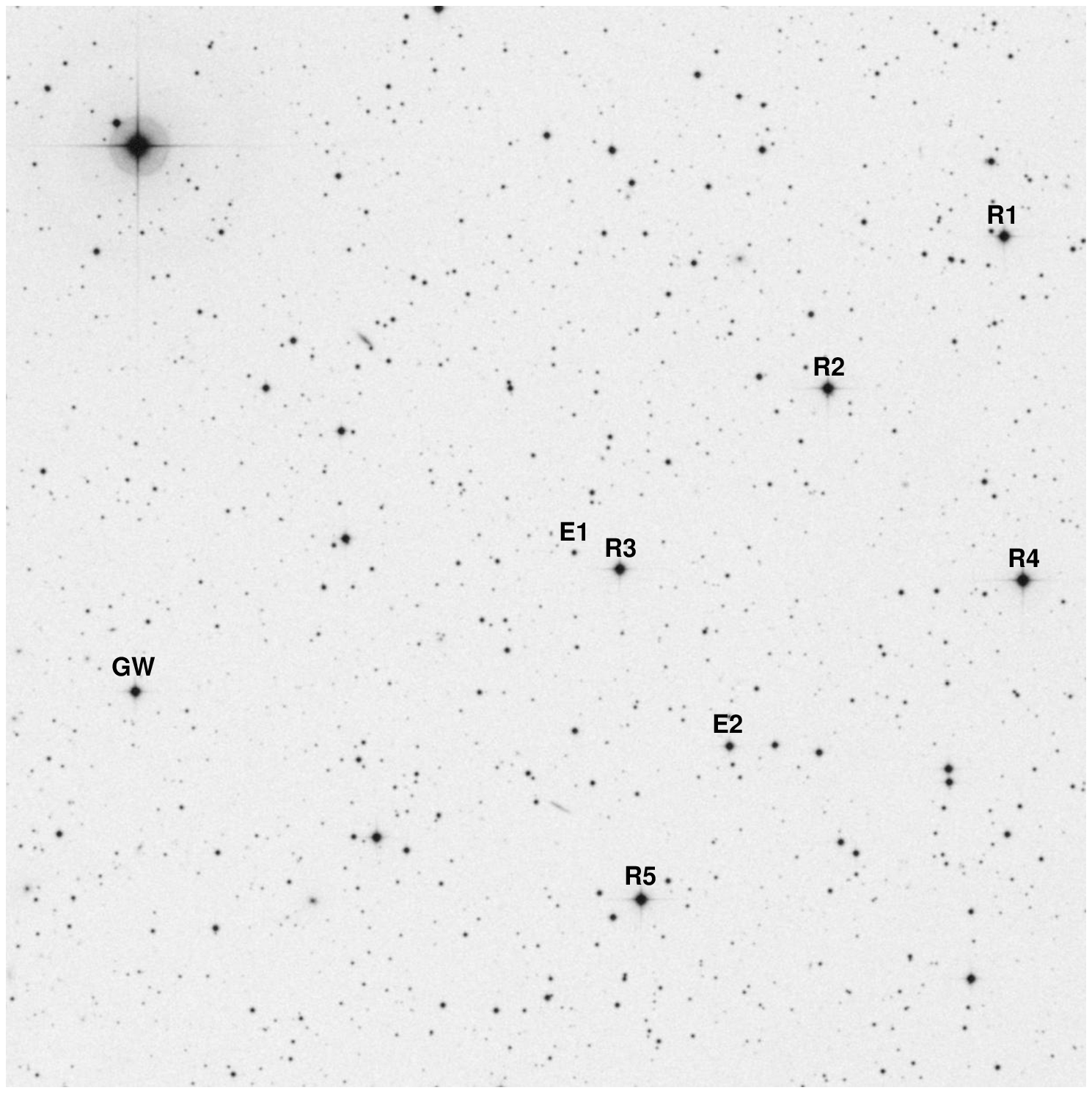}
 \caption{An observed CCD image (20$\arcmin$.5$\times$20\arcmin.5) of GW Cep and many nearby stars. E1 and E2 are 
 eclipsing binary systems named CzeV106 Cep (GSC 4502-1040) and CzeV079 Cep (GSC 4502-0138), respectively. 
 Monitoring numerous frames led us to choose R2 as a comparison star. North is up and east is to the left. }
 \label{Fig1}
\end{figure}

\begin{figure}
 \includegraphics[]{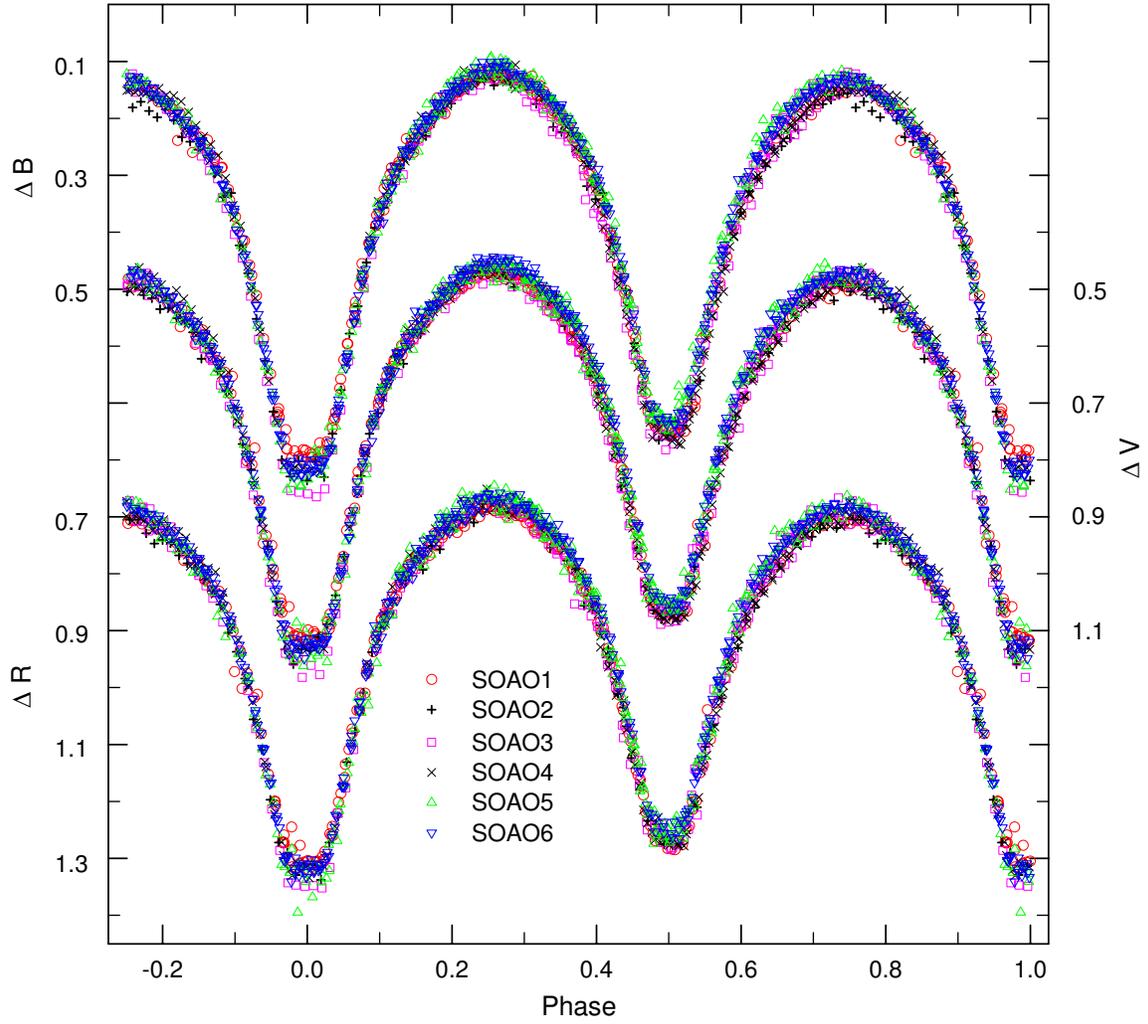}
 \caption{SOAO light curves of GW Cep in the $B$, $V$, and $R$ bandpasses defined by individual observations. }
 \label{Fig2}
\end{figure}

\begin{figure}
 \includegraphics[]{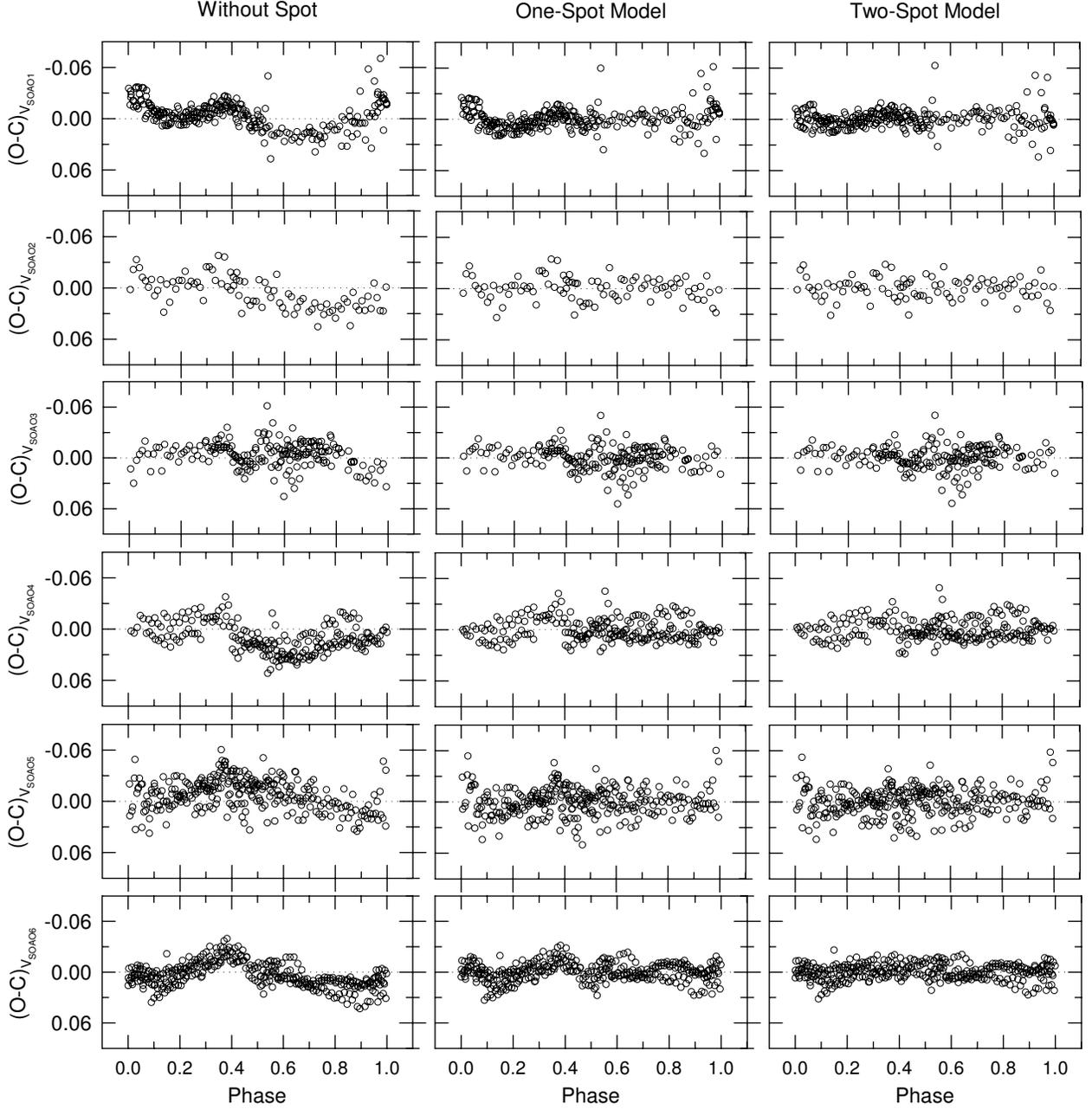}
 \caption{The left, middle, and right panels show the light residuals from the solutions without a spot, 
 with a one-spot model, and with a two-spot model listed in Tables 3 and 4, respectively. The data refer to 
 the $V$ bandpass. }
 \label{Fig3}
\end{figure}

\begin{figure}
 \includegraphics[]{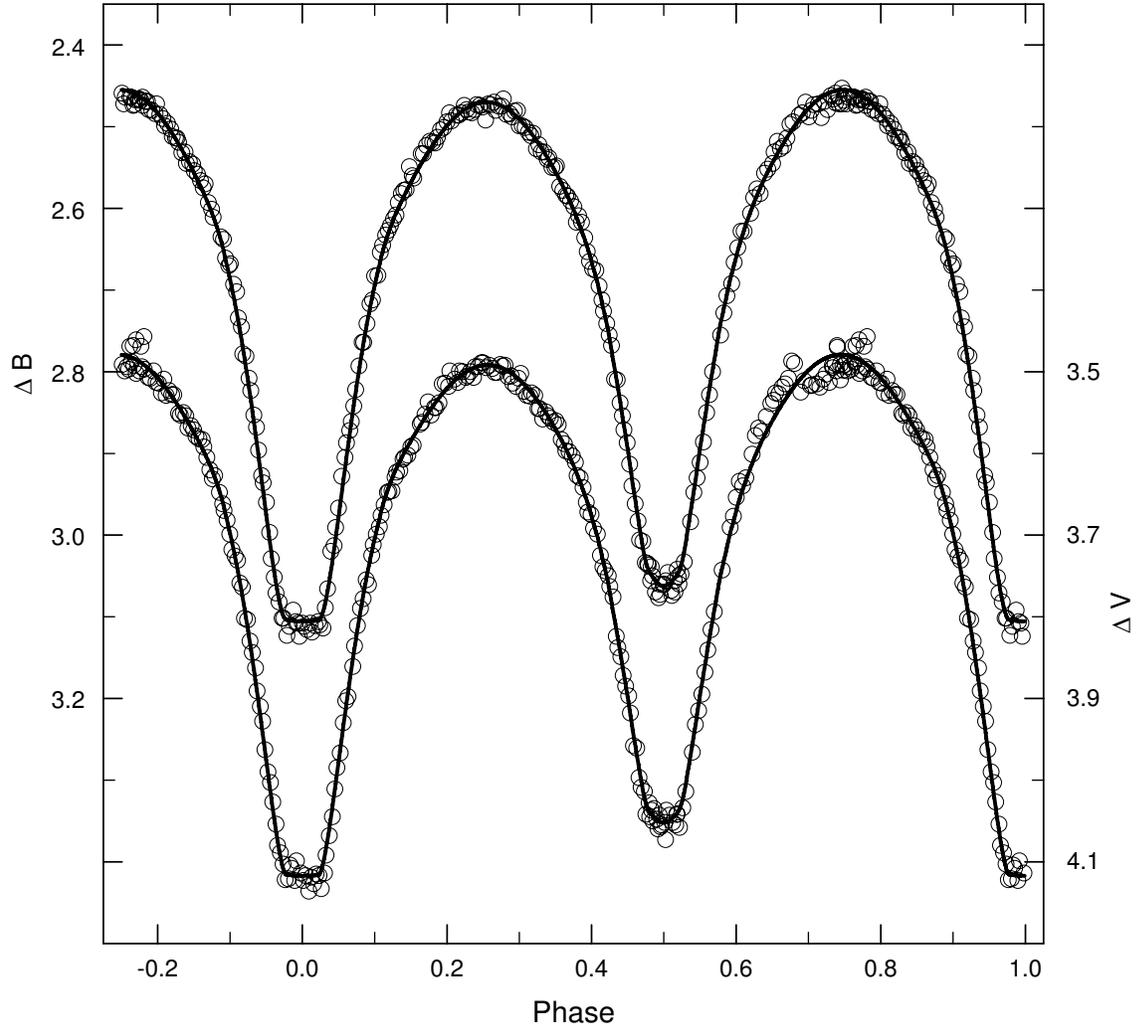}
 \caption{The $BV$ light curves of Hoffmann. The continuous curves represent the solutions obtained with 
 the model parameters listed in columns (2)-(3) of Table 5. }
 \label{Fig4}
\end{figure}

\begin{figure}
 \includegraphics[]{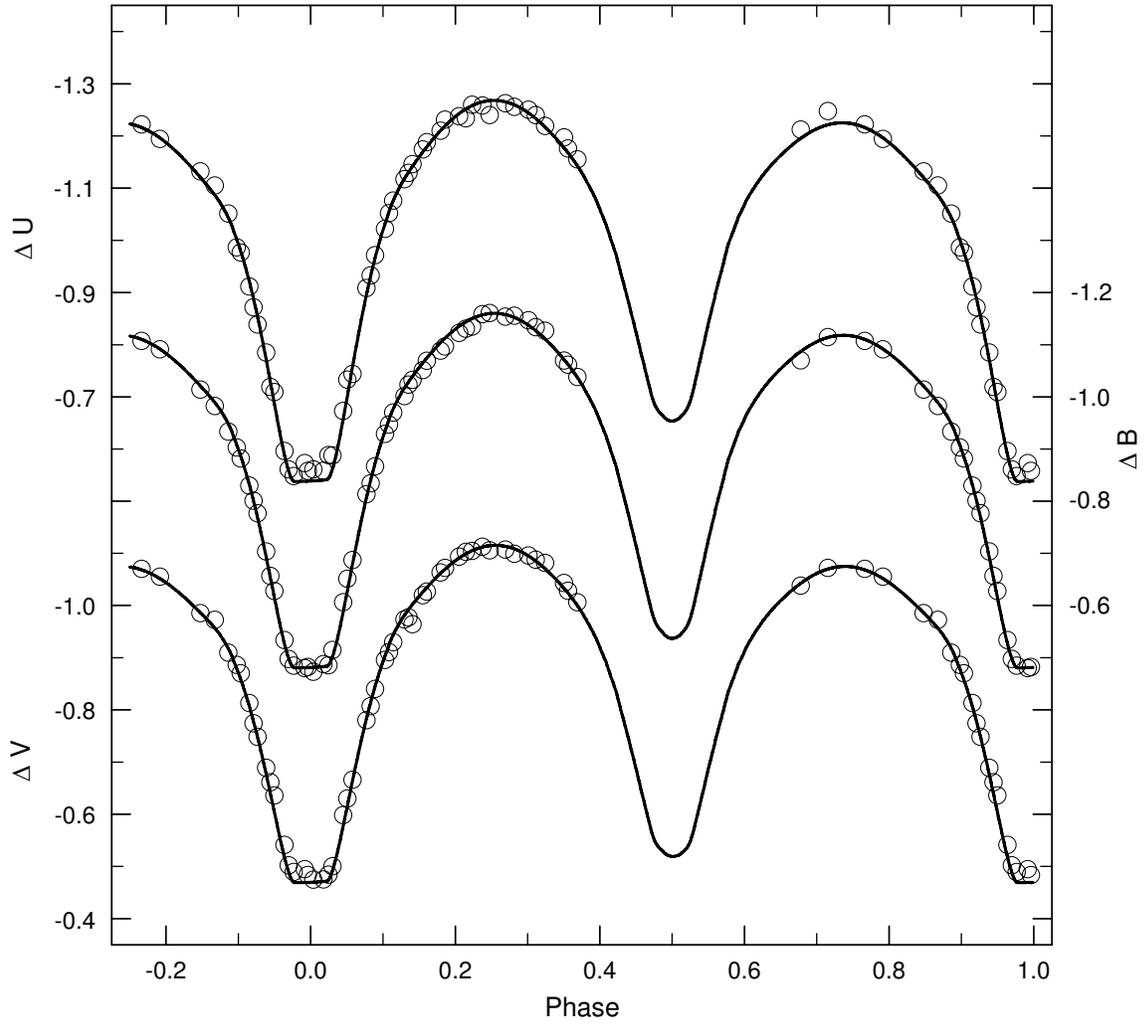}
 \caption{The $UBV$ light curves of Landolt. The continuous curves represent the solutions obtained with 
 the model parameters listed in columns (4)-(5) of Table 5. }
 \label{Fig5}
\end{figure}

\begin{figure}
 \includegraphics[]{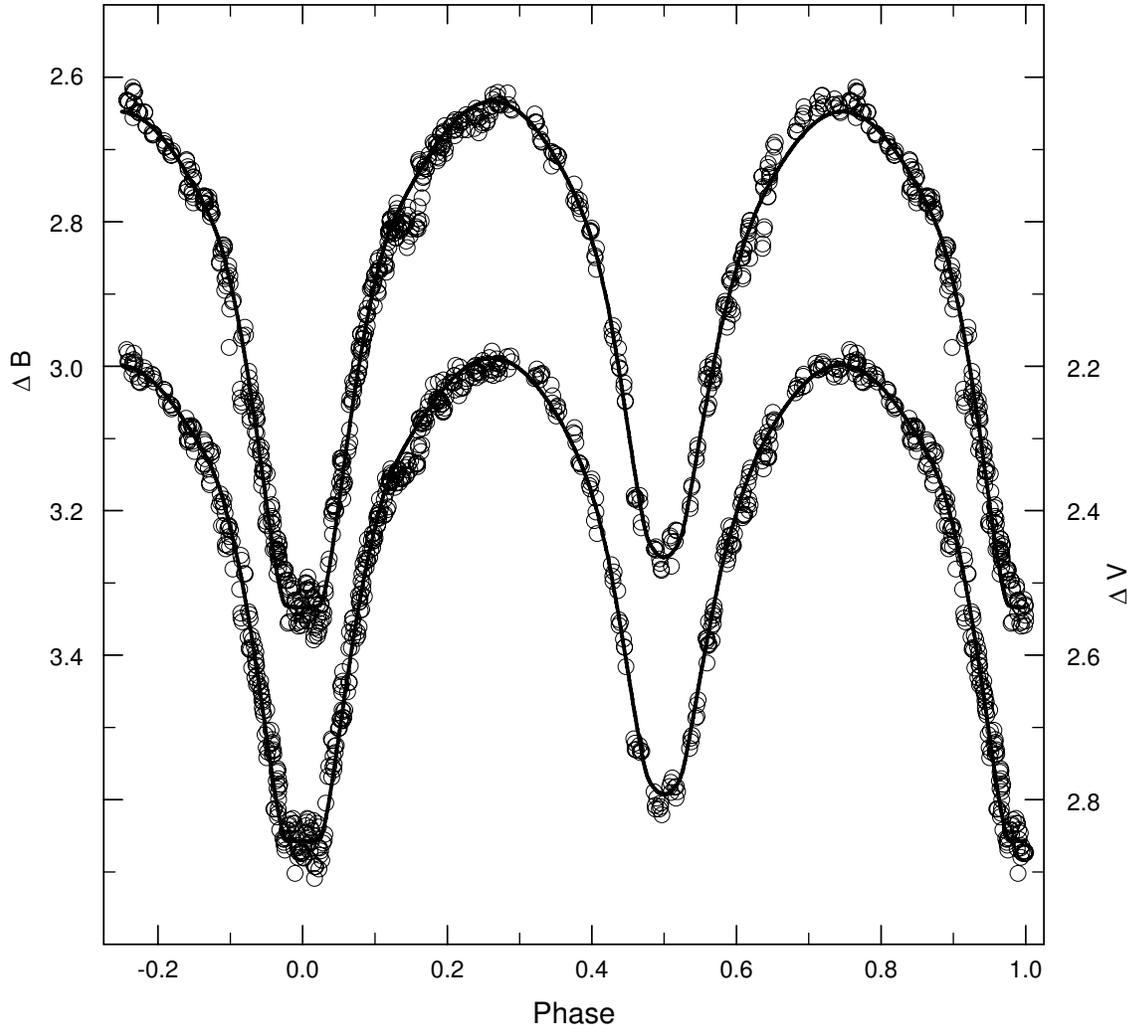}
 \caption{The $BV$ light curves of Pribulla et al. The continuous curves represent the solutions obtained with 
 the model parameters listed in columns (6)-(7) of Table 5. }
 \label{Fig6}
\end{figure}

\begin{figure}
 \includegraphics[]{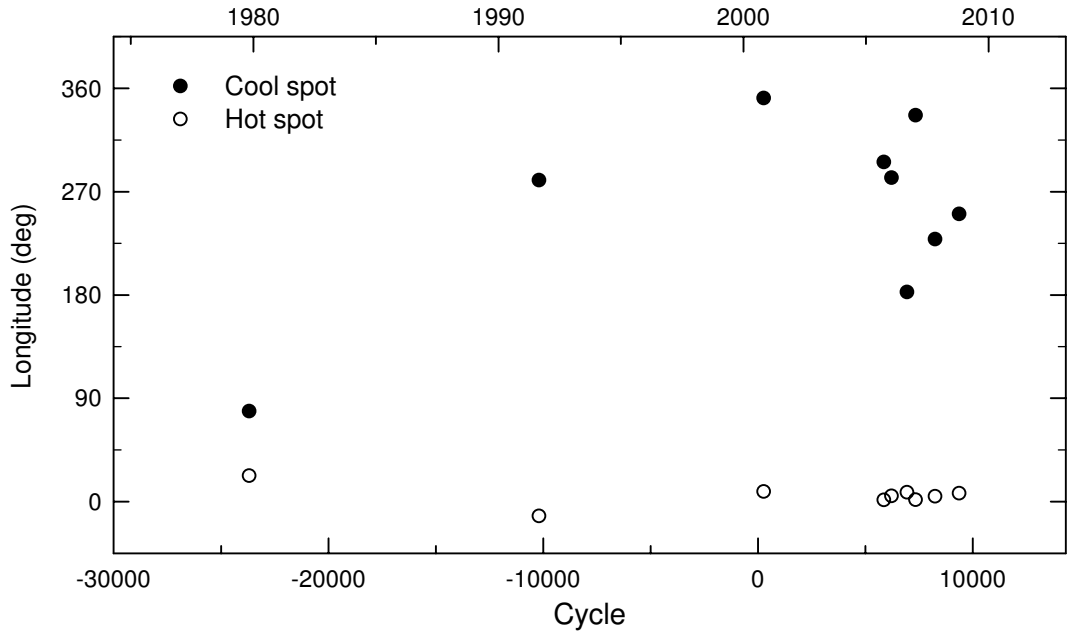}
 \caption{Changes of the cool- and hot-spot longitudes {\it versus} time from our two-spot model. }
 \label{Fig7}
\end{figure}

\begin{figure}
 \includegraphics[]{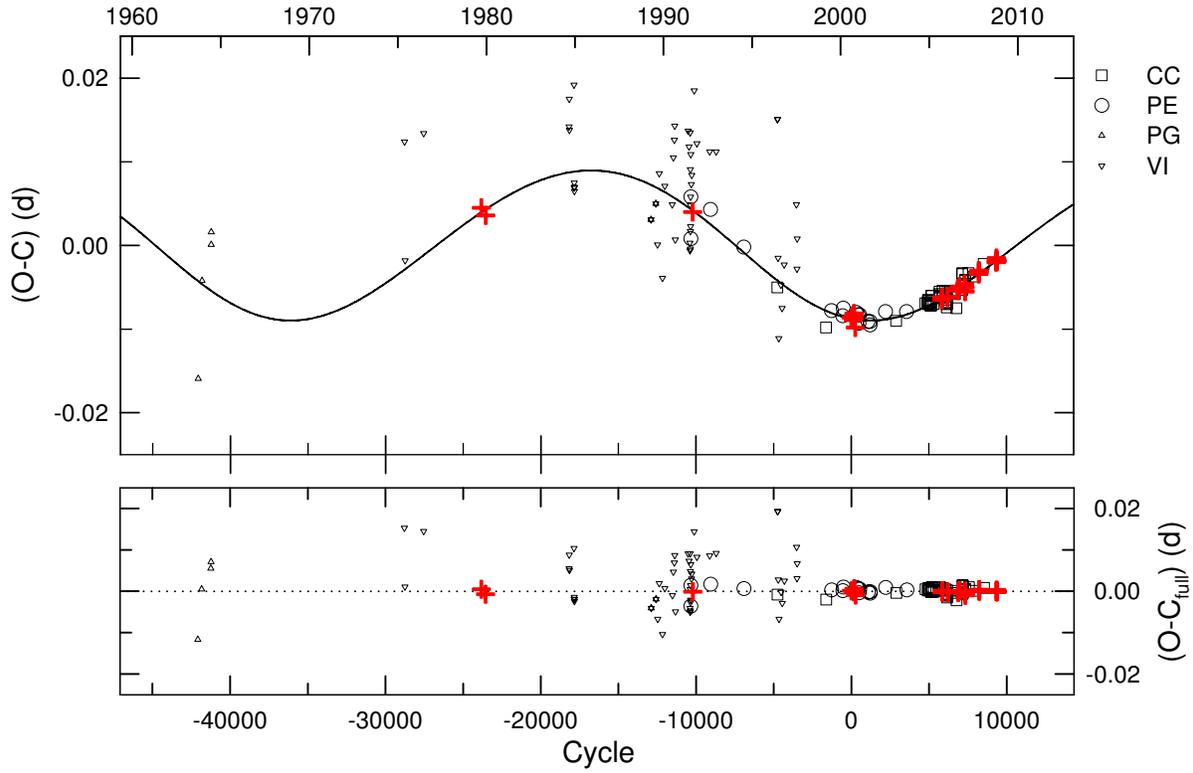}
 \caption{The $O$--$C$ diagram of GW Cep. In the upper panel, constructed with the linear terms in Table 7, 
 the continuous curve represents the LTT orbit. The residuals from the complete ephemeris are plotted 
 in the lower panel. Plus symbols refer to new minimum timings obtained with the W-D code. }
 \label{Fig8}
\end{figure}

\begin{figure}
 \includegraphics[]{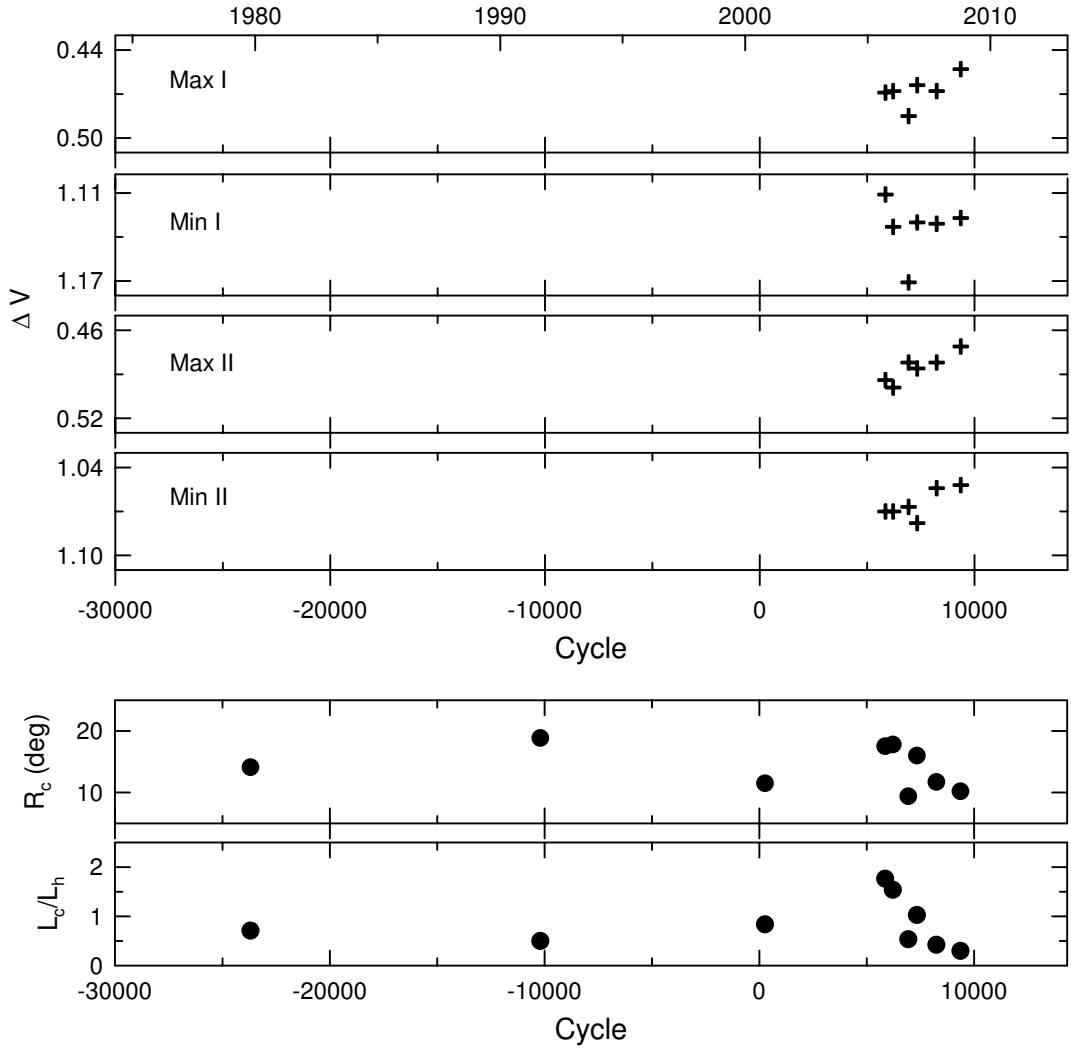}
 \caption{The four panels at the top display the $\Delta V$ variations for GW Cep at selected phases. The fifth 
 and sixth panels represent the changes of the radius ($\rm R_c$) of the cool spot and of the luminosity ratio 
 ($\rm L_c / L_h$) between the cool and hot spots, respectively, from our two-spot model. }
 \label{Fig9}
\end{figure}

\clearpage
\begin{deluxetable}{lcccc}
\tablewidth{0pt} 
\tablecaption{Summary of the SOAO observations of GW Cep.}
\tablehead{
\colhead{UT Date}  &  \colhead{Observing Interval}  &  \colhead{Filter}  &  \colhead{$N_{\rm obs}$}  &  \colhead{Note}  \\
                   &  (HJD+2,450,000)               &                    &                           &
}
\startdata
2005 Oct. 19, 20, 22, 23    &  3663.04$-$3667.06  &  $BVR$  &  796  &  SOAO1   \\  
2006 Feb. 11                &  3777.95$-$3778.29  &  $BVR$  &  271  &  SOAO2   \\  
2006 Sep. 10, 13, Oct. 18   &  3989.00$-$4027.07  &  $BVR$  &  576  &  SOAO3   \\  
2007 Jan. 29, Feb. 2, 6, 7  &  4129.92$-$4139.08  &  $BVR$  &  664  &  SOAO4   \\  
2007 Nov. 16, 17, 18, 24    &  4420.92$-$4429.38  &  $BVR$  &  892  &  SOAO5   \\  
2008 Nov. 5, 9, 10, 18, 19  &  4775.94$-$4790.21  &  $BVR$  &  1028 &  SOAO6   \\  
\enddata
\end{deluxetable}

\begin{deluxetable}{crcrcr}
\tablewidth{0pt} \tablecaption{CCD photometric observations of GW Cep.}
\tablehead{
\colhead{HJD} & \colhead{$\Delta B$} & \colhead{HJD} & \colhead{$\Delta V$} & \colhead{HJD} & \colhead{$\Delta R$} 
}
\startdata
2,453,663.04074 & 0.724  &  2,453,663.03811 & 0.992  &  2,453,663.03915 & 1.207   \\
2,453,663.04413 & 0.781  &  2,453,663.04197 & 1.068  &  2,453,663.04298 & 1.272   \\
2,453,663.04755 & 0.784  &  2,453,663.04536 & 1.117  &  2,453,663.04637 & 1.315   \\
2,453,663.05093 & 0.794  &  2,453,663.04876 & 1.110  &  2,453,663.04976 & 1.309   \\
2,453,663.05433 & 0.787  &  2,453,663.05215 & 1.117  &  2,453,663.05316 & 1.305   \\
2,453,663.05769 & 0.798  &  2,453,663.05556 & 1.118  &  2,453,663.05656 & 1.318   \\
2,453,663.06095 & 0.785  &  2,453,663.05887 & 1.112  &  2,453,663.05985 & 1.297   \\
2,453,663.06424 & 0.744  &  2,453,663.06214 & 1.085  &  2,453,663.06311 & 1.257   \\
2,453,663.06749 & 0.674  &  2,453,663.06541 & 1.056  &  2,453,663.06637 & 1.221   \\
2,453,663.07075 & 0.596  &  2,453,663.06866 & 0.979  &  2,453,663.06964 & 1.158   \\
\enddata
\tablecomments{This table is available in its entirety in machine-readable and Virtual Observatory (VO) forms 
in the online journal. A portion is shown here for guidance regarding its form and content.}
\end{deluxetable}

\begin{deluxetable}{lcc}
\tablewidth{0pt}
\tablecaption{GW Cep parameters obtained by fitting simultaneously all SOAO data.}
\tablehead{
\colhead{Parameter}      & \colhead{Primary}        & \colhead{Secondary}                
}                                                                                        
\startdata                                                                               
$T_0$ (HJD)              & \multicolumn{2}{c}{2,453,322.22167$\pm$0.00007}       \\ 
$P$ (d)                  & \multicolumn{2}{c}{0.31883212$\pm$0.00000002}         \\ 
$q$ (=m$\rm_s$/m$\rm_p$) & \multicolumn{2}{c}{2.590$\pm$0.008}                   \\ 
$i$ (deg)                & \multicolumn{2}{c}{85.4$\pm$0.2}                      \\ 
$T$ (K)                  & 6104$\pm$3                  & 5800                    \\ 
$\Omega$                 & 5.960$\pm$0.009             & 5.960                   \\ 
$f$ (\%)                 & \multicolumn{2}{c}{17.6}                              \\ 
$A$                      & 0.5                         & 0.5                     \\ 
$g$                      & 0.32                        & 0.32                    \\ 
$X$                      & 0.493                       & 0.506                   \\ 
$x_{B}$                  & 0.743$\pm$0.008             & 0.664$\pm$0.007         \\ 
$x_{V}$                  & 0.552$\pm$0.009             & 0.620$\pm$0.007         \\ 
$x_{R}$                  & 0.462$\pm$0.009             & 0.573$\pm$0.008         \\ 
$L/(L_1+L_2)_{B}$        & 0.357$\pm$0.001             & 0.643                   \\ 
$L/(L_1+L_2)_{V}$        & 0.353$\pm$0.001             & 0.647                   \\ 
$L/(L_1+L_2)_{R}$        & 0.347$\pm$0.001             & 0.653                   \\ 
$r$ (pole)               & 0.2881$\pm$0.0009           & 0.4435$\pm$0.0007       \\ 
$r$ (side)               & 0.3015$\pm$0.0011           & 0.4756$\pm$0.0009       \\ 
$r$ (back)               & 0.3404$\pm$0.0020           & 0.5052$\pm$0.0012       \\ 
$r$ (volume)$\rm ^a$     & 0.3119                      & 0.4763                  \\ 
$\Sigma W(O-C)^2$        & \multicolumn{2}{c}{0.0137}                            \\ 
\enddata
\tablenotetext{a}{Mean volume radius.}
\end{deluxetable}

\begin{deluxetable}{lcccccccccccc}
\tabletypesize{\tiny}
\tablewidth{0pt}
\rotate
\tablecaption{Spot and luminosity parameters for each SOAO dataset.}
\tablehead{
\colhead{Parameter}                     & \multicolumn{2}{c}{SOAO1}         & \multicolumn{2}{c}{SOAO2}         & \multicolumn{2}{c}{SOAO3}       & \multicolumn{2}{c}{SOAO4}         & \multicolumn{2}{c}{SOAO5}         & \multicolumn{2}{c}{SOAO6}          
}
\startdata                                                                                                                                                                                                                                                     
Mean Year                               & \multicolumn{2}{c}{2005.8}        & \multicolumn{2}{c}{2006.1}        & \multicolumn{2}{c}{2006.7}      & \multicolumn{2}{c}{2007.1}        & \multicolumn{2}{c}{2007.9}        & \multicolumn{2}{c}{2008.9}         \\
Mean Epoch                              & \multicolumn{2}{c}{5851}          & \multicolumn{2}{c}{6206}          & \multicolumn{2}{c}{6927}        & \multicolumn{2}{c}{7324}          & \multicolumn{2}{c}{8235}          & \multicolumn{2}{c}{9358}           \\[1.0mm] \hline \\[-2.0ex]
\multicolumn{13}{c}{One-Spot Model}                                                                                                                                                                                                                            \\[1.0mm] \hline \\[-2.0ex]
Phase shift$\rm ^a$                     & \multicolumn{2}{c}{$-$0.0004(1)}  & \multicolumn{2}{c}{$-$0.0006(3)}  & \multicolumn{2}{c}{$-$0.0009(2)}& \multicolumn{2}{c}{$-$0.0017(2)}  & \multicolumn{2}{c}{0.0006(2)}     & \multicolumn{2}{c}{0.0014(1)}      \\
Colatitude$_2$ (deg)                    & \multicolumn{2}{c}{72.2(1.1)}     & \multicolumn{2}{c}{72.7(5.6)}     & \multicolumn{2}{c}{79.7(3.2)}   & \multicolumn{2}{c}{79.7(6)}       & \multicolumn{2}{c}{76.8(3.4)}     & \multicolumn{2}{c}{74.8(6)}        \\
Longitude$_2$ (deg)                     & \multicolumn{2}{c}{292.5(1.1)}    & \multicolumn{2}{c}{273.2(5)}      & \multicolumn{2}{c}{181.8(1.3)}  & \multicolumn{2}{c}{327.5(3.6)}    & \multicolumn{2}{c}{227.3(2.2)}    & \multicolumn{2}{c}{240.0(1.1)}     \\
Radius$_2$ (deg)                        & \multicolumn{2}{c}{16.8(3)}       & \multicolumn{2}{c}{17.7(1.5)}     & \multicolumn{2}{c}{9.5(5)}      & \multicolumn{2}{c}{14.4(4)}       & \multicolumn{2}{c}{13.0(5)}       & \multicolumn{2}{c}{13.6(2)}        \\
$T$$\rm _{spot,2}$/$T$$\rm _{local,2}$  & \multicolumn{2}{c}{0.880(7)}      & \multicolumn{2}{c}{0.881(9)}      & \multicolumn{2}{c}{0.870(20)}   & \multicolumn{2}{c}{0.870(6)}      & \multicolumn{2}{c}{0.846(13)}     & \multicolumn{2}{c}{0.845(6)}       \\
$L_1/(L_1+L_2)_{B}$                     & \multicolumn{2}{c}{0.3567(3)}     & \multicolumn{2}{c}{0.3567(5)}     & \multicolumn{2}{c}{0.3567(4)}   & \multicolumn{2}{c}{0.3567(3)}     & \multicolumn{2}{c}{0.3567(3)}     & \multicolumn{2}{c}{0.3567(2)}      \\
$L_1/(L_1+L_2)_{V}$                     & \multicolumn{2}{c}{0.3530(3)}     & \multicolumn{2}{c}{0.3530(5)}     & \multicolumn{2}{c}{0.3529(4)}   & \multicolumn{2}{c}{0.3529(3)}     & \multicolumn{2}{c}{0.3529(3)}     & \multicolumn{2}{c}{0.3530(2)}      \\
$L_1/(L_1+L_2)_{R}$                     & \multicolumn{2}{c}{0.3470(3)}     & \multicolumn{2}{c}{0.3471(5)}     & \multicolumn{2}{c}{0.3471(4)}   & \multicolumn{2}{c}{0.3471(3)}     & \multicolumn{2}{c}{0.3471(3)}     & \multicolumn{2}{c}{0.3471(2)}      \\
$\Sigma W(O-C)^2$                       & \multicolumn{2}{c}{0.0100}        & \multicolumn{2}{c}{0.0103}        & \multicolumn{2}{c}{0.0116}      & \multicolumn{2}{c}{0.0096}        & \multicolumn{2}{c}{0.0132}        & \multicolumn{2}{c}{0.0085}         \\[1.0mm] \hline \\[-2.0ex]
\multicolumn{13}{c}{Two-Spot Model}                                                                                                                                                                                                                            \\[1.0mm] \hline \\[-2.0ex]
Phase shift$\rm ^a$                     & \multicolumn{2}{c}{$-$0.0008(1)}  & \multicolumn{2}{c}{$-$0.0016(3)}  & \multicolumn{2}{c}{$-$0.0013(2)}& \multicolumn{2}{c}{$-$0.0020(1)}  & \multicolumn{2}{c}{$-$0.0002(2)}  & \multicolumn{2}{c}{$-$0.0002(1)}   \\
Colatitude$_2$ (deg)                    &  75.6(8)      &  69.9(2.7)        &  75.6(0.9)    &  69.9(1.3)        &  80.0(1.1)    &  79.7(3.1)      &  77.9(9)      &  77.2(7)          &  77.2(4)      &  75.8(4.0)        &  76.2(3)      &  76.2(3.6)         \\
Longitude$_2$ (deg)                     &  1.4(1.0)     &  295.8(2.0)       &  4.9(1.1)     &  282.3(3.0)       &  8.1(1.0)     &  182.7(4.1)     &  1.6(5)       &  336.6(6)         &  4.7(1.2)     &  228.7(2.4)       &  7.3(4)       &  250.6(3.2)        \\
Radius$_2$ (deg)                        &  8.0(6)       &  17.5(5)          &  8.0(5)       &  17.8(6)          &  7.9(7)       &  9.4(5)         &  8.1(3)       &  16.0(2)          &  9.7(4)       &  11.7(1.0)        &  10.6(2)      &  10.2(3)           \\
$T$$\rm _{spot,2}$/$T$$\rm _{local,2}$  &  1.145(14)    &  0.892(6)         &  1.195(25)    &  0.892(8)         &  1.110(15)    &  0.870(20)      &  1.235(15)    &  0.884(5)         &  1.173(9)     &  0.861(14)        &  1.170(5)     &  0.879(9)          \\
$L_1/(L_1+L_2)_{B}$                     & \multicolumn{2}{c}{0.3567(3)}     & \multicolumn{2}{c}{0.3567(4)}     & \multicolumn{2}{c}{0.3567(3)}   & \multicolumn{2}{c}{0.3567(3)}     & \multicolumn{2}{c}{0.3567(3)}     & \multicolumn{2}{c}{0.3567(2)}      \\
$L_1/(L_1+L_2)_{V}$                     & \multicolumn{2}{c}{0.3530(3)}     & \multicolumn{2}{c}{0.3530(4)}     & \multicolumn{2}{c}{0.3530(4)}   & \multicolumn{2}{c}{0.3529(3)}     & \multicolumn{2}{c}{0.3529(3)}     & \multicolumn{2}{c}{0.3530(2)}      \\
$L_1/(L_1+L_2)_{R}$                     & \multicolumn{2}{c}{0.3471(3)}     & \multicolumn{2}{c}{0.3471(4)}     & \multicolumn{2}{c}{0.3471(4)}   & \multicolumn{2}{c}{0.3471(3)}     & \multicolumn{2}{c}{0.3471(3)}     & \multicolumn{2}{c}{0.3471(2)}      \\
$\Sigma W(O-C)^2$                       & \multicolumn{2}{c}{0.0098}        & \multicolumn{2}{c}{0.0095}        & \multicolumn{2}{c}{0.0114}      & \multicolumn{2}{c}{0.0091}        & \multicolumn{2}{c}{0.0126}        & \multicolumn{2}{c}{0.0072}         \\
\enddata
\tablenotetext{a}{From the data phased by the linear ephemeris of Table 3.}
\end{deluxetable}

\begin{deluxetable}{lcccccccc}
\tabletypesize{\small}
\tablewidth{0pt}
\tablecaption{Spot and luminosity parameters for historical light curves.}
\tablehead{
\colhead{Parameter}                     &  \multicolumn{2}{c}{Hoffmann (1982)}              &&  \multicolumn{2}{c}{Landolt (1992)}              &  \multicolumn{2}{c}{Pribulla et al. (2001)}         
}                                                                                                                                                
\startdata                                                                                                                                       
Mean Year                               & \multicolumn{2}{c}{1980.0}                        && \multicolumn{2}{c}{1991.8}                        &  \multicolumn{2}{c}{2000.9}                         \\
Mean Epoch                              & \multicolumn{2}{c}{$-$23695}                      && \multicolumn{2}{c}{$-$10208}                      &  \multicolumn{2}{c}{261}                            \\
$T_0$ (HJD)                             &  \multicolumn{2}{c}{2,444,200.48198$\pm$0.00006}  &&  \multicolumn{2}{c}{2,448,544.87119$\pm$0.00008}  &  \multicolumn{2}{c}{2,451,799.48430$\pm$0.00006}    \\
$P$ (day)                               &  \multicolumn{2}{c}{0.3188288$\pm$0.0000005}      &&  \multicolumn{2}{c}{0.318851025$\rm ^a$}          &  \multicolumn{2}{c}{0.3188282$\pm$0.0000003}        \\
Colatitude$_2$ (deg)                    &   67.0$\pm$4.7     &  71.8$\pm$1.1                &&   76.2$\pm$1.6     &  74.8$\pm$3.7                &   78.9$\pm$0.5     &  80.1$\pm$4.4                  \\
Longitude$_2$ (deg)                     &   22.5$\pm$2.2     &  78.8$\pm$1.8                &&   347.5$\pm$1.1    &  280.1$\pm$1.3               &   8.8$\pm$0.4      &  351.6$\pm$5.4                 \\
Radius$_2$ (deg)                        &   10.7$\pm$0.8     &  14.1$\pm$0.3                &&   12.8$\pm$0.5     &  18.8$\pm$0.6                &   7.1$\pm$0.3      &  11.5$\pm$2.3                  \\
$T$$\rm _{spot,2}$/$T$$\rm _{local,2}$  &   1.063$\pm$0.007  &  0.849$\pm$0.012             &&   1.153$\pm$0.017  &  0.798$\pm$0.018             &   1.280$\pm$0.011  &  0.962$\pm$0.018               \\
$L_1/(L_1+L_2)_{U}$                     &  \multicolumn{2}{c}{\dots}                        &&  \multicolumn{2}{c}{0.3693$\pm$0.0006}            &  \multicolumn{2}{c}{\dots}                          \\
$L_1/(L_1+L_2)_{B}$                     &  \multicolumn{2}{c}{0.3567$\pm$0.0003}            &&  \multicolumn{2}{c}{0.3526$\pm$0.0005}            &  \multicolumn{2}{c}{0.3567$\pm$0.0003}              \\
$L_1/(L_1+L_2)_{V}$                     &  \multicolumn{2}{c}{0.3530$\pm$0.0003}            &&  \multicolumn{2}{c}{0.3462$\pm$0.0005}            &  \multicolumn{2}{c}{0.3530$\pm$0.0003}              \\
\enddata
\tablenotetext{a}{From the ephemeris given by Landolt.}
\end{deluxetable}

\begin{deluxetable}{lcccccl}
\tablewidth{0pt}
\tablecaption{Minimum timings of GW Cep.}
\tablehead{
\colhead{Observed$\rm^{a}$} & \colhead{W-D$\rm^{a}$} & \colhead{Error$\rm^{b}$} & \colhead{Difference$\rm^{c}$} & \colhead{Filter} & \colhead{Min} & \colhead{References}}
\startdata
44,200.48205  &  44,200.48206  &  $\pm$0.00004  &  $-$0.00001  &  $BV$   &  I   &  Hoffmann (1982)          \\
44,289.27664  &  44,289.27556  &  $\pm$0.00015  &  $+$0.00108  &  $BV$   &  II  &  Hoffmann (1982)          \\
48,544.87103  &  48,544.87119  &  $\pm$0.00008  &  $-$0.00016  &  $UBV$  &  I   &  Landolt (1992)           \\
51,799.4839   &  51,799.48420  &  $\pm$0.00016  &  $-$0.00030  &  $BV$   &  I   &  Pribulla et al. (2001)   \\
51,845.3954   &  51,845.39552  &  $\pm$0.00011  &  $-$0.00012  &  $BV$   &  I   &  Pribulla et al. (2001)   \\
51,854.3240   &  51,854.32356  &  $\pm$0.00012  &  $+$0.00044  &  $BV$   &  I   &  Pribulla et al. (2001)   \\
51,858.3081   &  51,858.30860  &  $\pm$0.00010  &  $-$0.00050  &  $BV$   &  II  &  Pribulla et al. (2001)   \\
51,884.61130  &  51,884.61089  &  $\pm$0.00012  &  $+$0.00041  &  $BV$   &  I   &  Pribulla et al. (2001)   \\
53,663.05295  &  53,663.05284  &  $\pm$0.00009  &  $+$0.00011  &  $BVR$  &  I   &  This paper               \\
53,664.00969  &  53,664.00946  &  $\pm$0.00019  &  $+$0.00023  &  $BVR$  &  I   &  This paper               \\
53,664.16870  &  53,664.16890  &  $\pm$0.00015  &  $-$0.00020  &  $BVR$  &  II  &  This paper               \\
53,665.92277  &  53,665.92260  &  $\pm$0.00010  &  $+$0.00017  &  $BVR$  &  I   &  This paper               \\
53,666.08263  &  53,666.08188  &  $\pm$0.00006  &  $+$0.00075  &  $BVR$  &  II  &  This paper               \\
53,666.24113  &  53,666.24138  &  $\pm$0.00006  &  $-$0.00025  &  $BVR$  &  I   &  This paper               \\
53,777.99228  &  53,777.99175  &  $\pm$0.00010  &  $+$0.00053  &  $BVR$  &  II  &  This paper               \\
53,778.15077  &  53,778.15106  &  $\pm$0.00009  &  $-$0.00029  &  $BVR$  &  I   &  This paper               \\
53,989.05899  &  53,989.05873  &  $\pm$0.00010  &  $+$0.00026  &  $BVR$  &  II  &  This paper               \\
53,992.08747  &  53,992.08753  &  $\pm$0.00007  &  $-$0.00006  &  $BVR$  &  I   &  This paper               \\
53,992.24699  &  53,992.24680  &  $\pm$0.00011  &  $+$0.00019  &  $BVR$  &  II  &  This paper               \\
54,026.99936  &  54,026.99984  &  $\pm$0.00023  &  $-$0.00048  &  $BVR$  &  II  &  This paper               \\
54,129.98323  &  54,129.98218  &  $\pm$0.00008  &  $+$0.00105  &  $BVR$  &  II  &  This paper               \\
54,130.14180  &  54,130.14190  &  $\pm$0.00007  &  $-$0.00010  &  $BVR$  &  I   &  This paper               \\
54,134.12789  &  54,134.12692  &  $\pm$0.00013  &  $+$0.00097  &  $BVR$  &  II  &  This paper               \\
54,137.95373  &  54,137.95237  &  $\pm$0.00010  &  $+$0.00136  &  $BVR$  &  II  &  This paper               \\
54,421.07651  &  54,421.07667  &  $\pm$0.00014  &  $-$0.00016  &  $BVR$  &  II  &  This paper               \\
54,421.23582  &  54,421.23618  &  $\pm$0.00013  &  $-$0.00036  &  $BVR$  &  I   &  This paper               \\
54,422.35199  &  54,422.35188  &  $\pm$0.00013  &  $+$0.00011  &  $BVR$  &  II  &  This paper               \\
54,422.98958  &  54,422.98953  &  $\pm$0.00009  &  $+$0.00005  &  $BVR$  &  II  &  This paper               \\
54,423.14851  &  54,423.14906  &  $\pm$0.00012  &  $-$0.00055  &  $BVR$  &  I   &  This paper               \\
54,423.30857  &  54,423.30818  &  $\pm$0.00016  &  $+$0.00039  &  $BVR$  &  II  &  This paper               \\
54,776.09558  &  54,776.09595  &  $\pm$0.00009  &  $-$0.00037  &  $BVR$  &  I   &  This paper               \\
54,780.24062  &  54,780.24102  &  $\pm$0.00010  &  $-$0.00040  &  $BVR$  &  I   &  This paper               \\
54,781.19730  &  54,781.19767  &  $\pm$0.00004  &  $-$0.00037  &  $BVR$  &  I   &  This paper               \\
54,781.35713  &  54,781.35673  &  $\pm$0.00007  &  $+$0.00040  &  $BVR$  &  II  &  This paper               \\
54,789.16805  &  54,789.16824  &  $\pm$0.00006  &  $-$0.00019  &  $BVR$  &  I   &  This paper               \\
54,789.32798  &  54,789.32746  &  $\pm$0.00013  &  $+$0.00052  &  $BVR$  &  II  &  This paper               \\
54,789.96554  &  54,789.96515  &  $\pm$0.00005  &  $+$0.00039  &  $BVR$  &  II  &  This paper               \\
54,790.12440  &  54,790.12469  &  $\pm$0.00007  &  $-$0.00029  &  $BVR$  &  I   &  This paper               \\
\enddata
\tablenotetext{a}{HJD 2,400,000 is suppressed. }
\tablenotetext{b}{Uncertainties yielded by the W-D code. }
\tablenotetext{c}{Differences between columns (1) and (2). }
\end{deluxetable}

\begin{deluxetable}{lcc}
\tablewidth{0pt}
\tablecaption{Parameters for the LTT ephemeris of GW Cep.}
\tablehead{
\colhead{Parameter}            &  \colhead{Values}             &  \colhead{Unit}
}
\startdata
$T_0$                          &  2,451,799.49280$\pm$0.00026  &  HJD                   \\
$P$                            &  0.318830882$\pm$0.000000032  &  d                     \\
$a_{12}\sin i_{3}$             &  1.554$\pm$0.073              &  AU                    \\
$\omega$                       &  221.8$\pm$2.1                &  deg                   \\
$e$                            &  0.076$\pm$0.040              &                        \\
$n  $                          &  0.03021$\pm$0.00060          &  deg d$^{-1}$          \\
$T$                            &  2,450,716$\pm$72             &  HJD                   \\
$P_{3}$                        &  32.63$\pm$0.65               &  yr                    \\
$K$                            &  0.00896$\pm$0.00042          &  d                     \\
$f(M_{3})$                     &  0.00353$\pm$0.00018          &  $M_\odot$             \\
$M_3$ ($i_{3}$=90 deg)$\rm^a$  &  0.21                         &  $M_\odot$             \\
$M_3$ ($i_{3}$=60 deg)$\rm^a$  &  0.25                         &  $M_\odot$             \\
$M_3$ ($i_{3}$=30 deg)$\rm^a$  &  0.47                         &  $M_\odot$             \\
\enddata
\tablenotetext{a}{Masses of the hypothetical third body for different values of $i_{3}$.}
\end{deluxetable}

\begin{deluxetable}{lccc}
\tablewidth{0pt}
\tablecaption{Model parameters for possible magnetic activity of GW Cep.}
\tablehead{
\colhead{Parameter}       & \colhead{Primary}       & \colhead{Secondary}     & \colhead{Unit}
}
\startdata
$\Delta P$                &  0.1315                 &  0.1315                 &  s                     \\
$\Delta P/P$              &  $4.77\times10^{-6}$    &  $4.77\times10^{-6}$    &                        \\
$\Delta Q$                &  ${9.84\times10^{48}}$  &  ${2.67\times10^{49}}$  &  g cm$^2$              \\
$\Delta J$                &  ${6.61\times10^{46}}$  &  ${1.27\times10^{47}}$  &  g cm$^{2}$ s$^{-1}$   \\
$I_{\rm s}$               &  ${1.13\times10^{53}}$  &  ${7.51\times10^{53}}$  &  g cm$^{2}$            \\
$\Delta \Omega$           &  ${5.87\times10^{-7}}$  &  ${1.69\times10^{-7}}$  &  s$^{-1}$              \\
$\Delta \Omega / \Omega$  &  ${2.57\times10^{-3}}$  &  ${7.40\times10^{-4}}$  &                        \\
$\Delta E$                &  ${7.76\times10^{40}}$  &  ${4.28\times10^{40}}$  &  erg                   \\
$\Delta L_{\rm rms}$      &  ${2.37\times10^{32}}$  &  ${1.31\times10^{32}}$  &  erg s$^{-1}$          \\
                          &  0.061                  &  0.033                  &  $L_\odot$             \\
                          &  0.108                  &  0.030                  &  $L_{1,2}$             \\
$\Delta m_{\rm rms}$      &  $\pm$0.039             &  $\pm$0.021             &  mag                   \\
$B$                       &  11.2                   &  7.9                    &  kG                    \\
\enddata
\end{deluxetable}

\end{document}